\documentstyle[11pt]{article}
\begin{document}
\begin{titlepage}
\begin{flushright}
SUSSEX-AST 94/3-1 \\
IMPERIAL/TP/93--94/21 \\
FERMILAB-Pub-94/048-A \\
astro-ph/9403005 \\
(Submitted to {\bf Physical Review D})\\
\end{flushright}
\begin{center}
\LARGE
{\bf Texture-Induced Microwave Background Anisotropies}\\
\vspace{.6cm}
\normalsize
\large{Julian Borrill$^{1,2}$, Edmund J.~Copeland$^1$, Andrew R.~Liddle$^1$,\\
\vspace*{4pt}
Albert Stebbins$^3$ and Shoba Veeraraghavan$^{4,5}$} \\
\normalsize
\vspace{.4 cm}
{\em $^1$School of Mathematical and Physical Sciences, \\
University of Sussex, \\ Falmer, Brighton BN1 9QH,~~~U.~K.}\\
\vspace{.3 cm}
{\em $^2$Blackett Laboratory$^*$,\\
Imperial College of Science and Technology,\\
Prince Consort Road, London SW7 2BZ,~~~U.~K.}\\
\vspace{.3 cm}
{\em $^3$NASA/Fermilab Astrophysics Center,\\ Fermi National Accelerator
Laboratory,\\ Batavia, Illinois 60510,~~~U.~S.~A.}\\
\vspace{.3 cm}
{\em $^4$NASA Goddard Space Flight Center$^*$,\\
Code 685, Greenbelt, MD 20771,~~~U.~S.~A.}\\
\vspace{.3 cm}
{\em $^5$Steward Observatory,\\ University of Arizona, \\
Tucson, Arizona 85721,~~~U.~S.~A.}\\
\vspace{.3 cm}
\end{center}
\baselineskip=24pt
\begin{abstract}
\noindent
We use numerical simulations to calculate the cosmic microwave
background anisotropy induced by the evolution of a global texture
field, with special emphasis on individual textures.  Both spherically
symmetric and general configurations are analysed, and in the latter
case we consider field configurations which exhibit unwinding events
and also ones which do not. We compare the results given by evolving
the field numerically under both the expanded core (XCORE) and
non-linear sigma model (NLSM) approximations with the analytic
predictions of the NLSM exact solution for a spherically symmetric
self-similar (SSSS) unwinding. We find that the random unwinding
configuration spots' typical peak height is 60--75\% and angular size
typically only 10\% of those of the SSSS unwinding, and that random
configurations without an unwinding event nonetheless may generate
indistinguishable hot and cold spots. The influence of these results
on analytic estimates of texture induced microwave anisotropies is
examined, and comparison made with other numerical work.
\end{abstract}
\vspace{-.3cm}
\begin{center}
{\small PACS numbers: \hspace{0.5cm} 98.80.Cq, 98.70.Vc}\\
\end{center}
\vspace{-.4cm}
{\small $^*$Present address.}
\end{titlepage}


\section{Introduction}

The main rival to the inflationary cosmology \cite{INFL} in explaining
the origin of structure in the universe remains theories based on
inhomogeneities formed at phase transitions. These inhomogeneities may
be seeded by topologically stabilised objects such as cosmic strings
\cite{STRING}, by the correlation of topologically trivial variations
in the local direction of symmetry breaking, as with nontopological
texture \cite{TEX}, or by a combination of the two, as with global
strings \cite{GS}, monopoles \cite{BR} or textures \cite{TEX}. A
proper assessment of the viability of such models requires a detailed
understanding of their influence on the isotropy of the cosmic
microwave background (CMB). In this paper we shall study some aspects
of the CMB anisotropy produced by global textures.

Several authors have made studies of the predictions of the texture
model for large scale structure \cite{TEXLSS} and CMB anisotropies
\cite{TEXMTS,DS,BR2,PST,CPT}. The CMB predictions of textures can be
compared with observations by the Differential Microwave Radiometer
(DMR) on the Cosmic Background Explorer (COBE) satellite \cite{COBE}
to fix the one-parameter normalisation of the theory; however, the
amplitude of density inhomogeneities then seems very small when
normalized to COBE \cite{BR2,PST}. Most CMB anisotropy calculations
employ numerical simulations of the texture field evolution over a
large volume of space, in some cases including the entire observable
universe, and calculate the field's stress-energy tensor at each point
at each time-step. Another approach involves calculating the
anisotropy due to a single texture analytically and summing the result
over an ensemble of textures with the appropriate number density and
distribution.

A feature of defect field evolution peculiar to texture is that the
field may order itself in two distinct ways. One mechanism is simply
by smoothing its spatial gradients, a process which involves no
topology change. The alternative is in a process known as `unwinding'.
During unwinding the field gradients near a point increase to such an
extent that it becomes energetically possible for the field to pull
itself off the vacuum manifold and over the symmetric-vacuum energy
barrier, changing the local topological charge by unity. The only
known non-trivial analytic solution of the texture field equations
describes such an unwinding in the Non Linear Sigma Model (NLSM)
approximation \cite{TEXMTS}. It is therefore standard to utilise this
SSSS exact unwinding solution when describing textures analytically,
including in the determination of CMB anisotropy detailed above.
Despite breaking down at the unwinding event itself, where a
singularity occurs and one must patch the ingoing to the outgoing
solution, in general the NLSM approximation is an excellent one;
however spherical symmetry and self-similarity are not good
approximations to realistic configurations. We are therefore interested
in how more general configurations compare with the SSSS case.

Recently, three of the present authors have published a two part
numerical study of individual textures throughout which the NLSM
approximation is lifted and the full field equations, including the
potential term for the texture field, are used, giving non-singular
evolution through the unwinding event. In the first part of the study
\cite{BCL1} the assumption of spherical symmetry is retained but not
that of self-similarity. In the second part \cite{BCL2} the
spherically symmetric assumption is also dropped, allowing the study
of texture unwinding events arising from randomly generated initial
field configurations.  One of the conclusions of the study is that the
properties of the SSSS solution are not characteristic of those of
more realistic random configurations (see also Ref.~\cite{NSY}). There
are two aspects to this.  The first, a problem with all spherically
symmetric configurations, is that spherical symmetry sets up very
large scale correlations in the field, so that sections of the texture
which are causally separated behave in a coherent way. The second is
associated with the self-similarity; the initial field conditions also
include a coherent velocity to keep it in the self-similar state,
which provides a coherent `push' on the texture towards unwinding.
In the SSSS case this results in the texture unwinding as fast as
is causally possible. Since the size of the CMB anisotropy is
associated with the rate of change of the metric perturbations induced
by the evolving field we might expect the typical anisotropy due to a
more realistic texture configuration to have a less pronounced
signature than that of the SSSS solution. In this paper we first
calculate the anisotropy induced by a spherically symmetric but
non-self-similar field configuration, and then those induced by
general, randomly generated, configurations which have no particular
symmetries or coherences. We thereby address the issue of how the CMB
anisotropy patterns from realistic texture unwinding and non-unwinding
configurations compare with that of the SSSS unwinding.

We wish to make a comparative study of CMB anisotropy produced by
localized excitations of the texture field. Various processes can be
important in determining the CMB anisotropy; e.g. Thomson scattering
may or may not be important depending on whether the excitation is in
front of or on the surface of last scattering, cosmological expansion
can be more or less important depending on whether the coherence
length of the texture excitation is large or small compared to the
cosmological horizon, and finally various geometrical and physical
effects are more or less important depending whether the excitation
subtends a large or small angle on the sky. Since we are interested in
a comparison of the different texture configurations and since the
processes mentioned probably affect the different texture
configurations in a similar way we feel it is reasonable to ignore all
of these effects and to compare the CMB anisotropy for the texture
excitations in the simplest possible environment.  Here we will
compare temperature patterns for texture excitations in Minkowski
space. This is the appropriate limit if the texture excitations
subtend a small angle, are well in front of the surface of last
scattering and are much smaller than the horizon when we see them.
This is independent of the redshift at which we see them.

An important consideration in computing any effect of textures in a
cosmological setting involves determining appropriate initial
conditions for both the texture fields and the matter fields. Energy
and momentum are always locally conserved, so that the energy
overdensity corresponding to the presence of a topological defect is
locally compensated by an energy underdensity in the other fields
\cite{VS90}. Na\"{\i}vely we might assume that compensation, by
restoring the initial homogeneity of the total energy density, would
automatically reduce the CMB anisotropy generated by a particular
defect field configuration. However, as two of the present authors
have shown \cite{VS92SV93a}, in many cases the compensating
perturbations will increase and not decrease both the density
perturbation and CMB anisotropy. Whether or not compensation will
decrease or increase the CMB anisotropy for realistic texture
configurations is not clear; however, the analytic work just cited
suggests that we should expect no strong cancellation of CMB
anisotropies due to compensation.

\section{Texture Field Evolution}

The central problem in numerical studies of texture evolution is that
the ratio of the two fundamental length scales --- the macrophysical
field correlation length and the microphysical field width --- is so
large that it is impossible to resolve them on the simulation lattice
simultaneously.  We therefore have to adopt one of two approximations,
either massively enlarging the field width (the XCORE approximation)
or shrinking it to zero (the NLSM approximation). If these results are
to be taken seriously, it is crucial that these two different
approximation methods give comparable results, and we shall see that
this is indeed the case. A more detailed comparison and critique of
these two approaches is given separately by one of us \cite{B94}.

We work throughout in flat space with the texture field normalised
such that the vacuum manifold is the 3-sphere $|\Phi| = 1$, so the
potential is given by $V(\Phi) = V_0 (\Phi^2 - 1)^2$.  Discretising
the equations of motion in both space and time then yields equations
for $\Phi_{i,n}$ (where i-indices are spatial and n-indices temporal),
giving the field at points separated by the grid-spacing $\delta x$
and by the time step $\delta t$.

\subsection{The XCORE Approximation}

In flat space the equation of motion of the texture field $\Phi$ is
\begin{equation}
\ddot{\Phi} - \nabla^2 \Phi = - 4 V_{\rm o} (\Phi^2 - 1) \Phi
\end{equation}
Given the field at some time $t$ and its velocity at time $(t -\delta t)/2$
these may now be discretised to second order using a standard staggered
leapfrog approach \cite{PFTV}, giving
\begin{eqnarray}
\dot{\Phi}_{i,n+1/2} & = & \dot{\Phi}_{i,n-1/2} + \left(\nabla^2 \Phi_{i,n}
	- 4 V_{\rm o} (\Phi^{2}_{i,n} - 1) \Phi_{i,n} \right) \delta t
\nonumber \\
\Phi_{i,n+1} & = & \Phi_{i,n} + \dot{\Phi}_{i,n+1/2} \delta t
\end{eqnarray}
In order to be able to resolve the texture's core the field width must
be unrealistically large, equivalent to making the mass of the radial
mode unrealistically small. This approach was introduced for domain
wall simulations in Ref.~\cite{PRS}, and for texture simulations in
Ref.~\cite{STPR}. Care must be taken to ensure that this approximation
does not qualitatively affect the field's evolution, with the degree
of the approximation being characterised by the ratio of the field
width to its correlation length. We therefore work with an oversized
field width
\begin{equation}
W_{\rm o} \equiv m_{\Phi}^{-1} = 1/\sqrt{8 V_{\rm o}} = 0.25 \; \delta x
\end{equation}
on a grid with $96^3$ points. We employ the catalogue of unwinding and
non-unwinding random initial configurations developed in Ref.
\cite{BCL2}; the initial configurations are set by randomly assigning
field values on the manifold on a $3 \times 3$ grid, and interpolating
between them (accounting for periodic boundary conditions) onto the
$96^3$ grid. The initial velocities are set to zero. A crude measure
of the correlation length is therefore the separation of the
uncorrelated points, and this value is large enough compared to the
field width to be known not to qualitatively distort the field's
evolution \cite{B93}.  Having 27 independent correlation volumes of the
field gives a reasonable chance, about 4\%, of obtaining a texture
unwinding in a simulation.

In so far as we can take the distance of 32 grid
points between the randomly selected points to represent the
correlation length, this provides an estimate of the initial `horizon
radius' in the simulations\footnote{We are treading in dangerous waters here by
mentioning horizon sizes whilst carrying out simulations in flat
space. In expanding universes the horizon size is closely connected to
the Hubble length, whereas here we intend the term to indicate the
distance light can have travelled between some conceptual initial time
when the correlation length approached zero and the time the
simulation actually started, motivated by the knowledge that the
texture field correlations grow at or close to the speed of light.},
which we note is considerably larger in grid units than in other
simulations, e.~g.~Refs.~\cite{PST,CPT}. The simulations are run for a
maximum time of 48 grid units (though of course the actual number of
time steps is much greater); further forward evolution endangers
problems with the periodic boundary conditions.

For comparison we also consider a spherically symmetric
non-self-similar initial configuration of the form
\begin{eqnarray}
\Phi(r, \theta, \psi) & = & (\cos \chi(r), \sin \chi(r) \cos \theta, \sin
\chi(r) \sin \theta \cos \psi, \sin \chi(r) \sin \theta \sin \psi) \\
\dot{\Phi} & = & 0
\end{eqnarray}
with
\begin{equation}
\label{SS}
\chi(r) = \pi (1 - e^{-r/r_{\rm o}})
\end{equation}
and take $r_{\rm o} = 18 \; \delta x$ to satisfy the lattice resolution
constraints.

\subsection{The NLSM Approximation}

In the NLSM approximation, the flat space equations of motion of the
texture field become
\begin{equation}
\ddot{\Phi} - \nabla^2 \Phi = - (\dot{\Phi}^2 - \nabla \Phi^2) \Phi
\end{equation}
Given the field at times $t$ and $t - \delta t$ the equations
of motion may be discretised to second order to give
\begin{equation}
\Phi_{i,n+1} = 2 \Phi_{i,n} - \Phi_{i,n-1} + \nabla^2 \Phi_{i,n} \delta
t^2 - (\dot{\Phi}_{i,n}^2 - \nabla \Phi_{i,n}^2)\delta t^2
\Phi_{i,n}
\end{equation}
Following Ref.~\cite{PST}, we write this in the form
\begin{equation}
\label{NLSMEVO}
\Phi_{i,n+1} = \delta \Phi_{i,n} + \lambda \Phi_{i,n}
\end{equation}
with
\begin{eqnarray}
\delta \Phi_{i,n} & = & \Phi_{i,n} - \Phi_{i,n-1} + \nabla^2
\Phi_{i,n} \delta t^2 \nonumber\\
\lambda & = & 1 - (\dot{\Phi}_{i,n}^2 - \nabla \Phi_{i,n}^2)\delta
t^2
\end{eqnarray}
we can use the constraint $|\Phi_{i,n+1}| = 1$ to solve for $\lambda$,
giving
\begin{equation}
\label{UW}
\lambda = - \delta \Phi_{i,n}.\Phi_{i,n} \pm \sqrt{1 - \delta
\Phi_{i,n}^{2}  + (\Phi_{i,n}.\delta \Phi_{i,n})^2}
\end{equation}

Since $\delta \Phi_{i,n}$ is small
\begin{equation}
\Phi_{i,n+1} \sim \left\{
\begin{array}{ll}
+ \Phi_{i,n} & $taking the positive root in$ \; \lambda \nonumber \\
- \Phi_{i,n} & $taking the negative root in$ \; \lambda
\end{array} \right.
\end{equation}
The choice of the negative root in the evaluation of $\lambda$ is
therefore equivalent to explicitly introducing an unwinding event at
grid position $i$ at time-step $n$. Since the NLSM approximation breaks
down at unwindings, this provides a means of re-introducing them.
However although this explicit re-introduction may be necessary for
spherically symmetric configurations, whose unwinding sites can be
predetermined and made to lie exactly on a gridpoint, for general
field configurations unwindings occur implicitly, off the grid, and we
should always take the positive root in $\lambda$. Indeed to do
otherwise \cite{PST,CPT} is not only unnecessary but also incorrect
since under the NLSM approximation {\em only} at the unwinding site
itself should the field ever leave the vacuum manifold \cite{B94}.

We can now take the same set of initial field configurations as in the
XCORE runs, with identical simulation parameters, but now evolved via
Eq.~(\ref{NLSMEVO}). For the spherically symmetric configuration,
the unwinding occurs at a grid point and is explicitly introduced by
taking the negative root in Eq.~(\ref{UW}) when the field within
one grid spacing of the unwinding site covers more than half of the
vacuum manifold.

\section{The CMB Anisotropy}

\subsection{Analytic Formalism}

The evolving texture field will produce an inhomogeneous time-varying
gravitational field. Photons travelling along different trajectories
will gain or lose different amounts of energy in the time varying
field which will thus result in different shifts in temperature of the
CMB photons in different directions on the sky. For the weak fields
produced by textures the temperature shifts are given by the
Sachs-Wolfe integral \cite{SW}. The Sachs-Wolfe integral may be
re-expressed as an integral over the stress-energy distribution which
produces the gravitational fields. This last integral was examined in
Ref.~\cite{S88} for sources which are far from the observer and
subtend small angles in flat space, and re-examined by Hindmarsh
\cite{H93} in a rather clever and simple way which also
applies to gravitational lensing. Extensions to large angles are given
in Ref.~\cite{SV93b} for flat space and in Ref.~\cite{VS93} for a
matter-dominated FRW cosmology. Here we use the small-angle
large-distance approximation in flat space. In this case the
temperature pattern we see depends only on the stress-energy on the
past light cone of the observer.

In the small-angle large-distance approximation the photons we see at
one instant were approximately in a plane when they passed by the
object we are viewing. We may thus label the temperature pattern by a
2-dimensional vector, ${\bf x}_\perp$, which is perpendicular to the
direction ${\bf{\hat{n}}}$ in which we are looking. The congruence of
photons may thus be approximated as ${\bf x}_\gamma(t) = {\bf x}_\perp
- {\bf{\hat{n}}}(t - t_{i})$ where $t_{i}$ is the impact time, so this
plane passes by the observers at the time of observation. With this
notation the temperature pattern is given by
\cite{SV93b,H93}
\begin{equation}
\nabla^2_{\perp} \frac{\Delta T}{T}({\bf x}_\perp)
	= - 8\pi G \nabla_{\perp} \cdot{\bf U}({\bf x}_\perp)
\label{eAS1}
\end{equation}
where the 2-dimensional vector ${\bf U}$ is ($i,j,k,\ldots$ are spatial
indices)
\begin{equation}
{\bf U}_{i}({\bf x}_\perp) = -\left(\delta_i^j - {\hat n}_i {\hat n}^j
	\right) \int_{- \infty}^{\infty} \left( \Theta_{0j}(t,{\bf x}_\gamma
	(t)) - \hat{n}^k \Theta_{jk}(t,{\bf x}_\gamma(t))
	\right)\,{\rm d}t
\label{eAS2}
\end{equation}
and $\Theta_{\mu\nu}(t,{\bf x})$ is the stress-energy tensor.
Note that Eq.~(\ref{eAS1}) only determines the temperature pattern up
to an arbitrary function satisfying $\nabla_\perp^2 \phi = 0$. However setting
zero boundary conditions at infinity determines the function uniquely, while
the
periodic boundary conditions that we shall use determine $\phi$ up to a
constant. This constant is chosen so that the mean anisotropy is zero.
The texture stress-energy tensor is
\begin{equation}
\Theta_{\alpha\beta}=\partial_{\alpha}\Phi\partial_{\beta}\Phi-
	g_{\alpha\beta}{\cal L}
\end{equation}
and taking the line of sight to be ${\bf\hat{n}}=(-1,0,0)$,
Eq.~(\ref{eAS2}) becomes
\begin{equation}
{\rm U}_i({\bf x}_{\gamma}^{\perp}) = - \int_{- \infty}^{\infty}
	(\partial_0 \Phi+\partial_1 \Phi)\partial_i \Phi\;{\rm d}t
\label{eAS3}
\end{equation}
for $i=2,\,3$.  Given a field configuration and its initial
distribution, Eqs.~(\ref{eAS2}) and (\ref{eAS3}) determine the
anisotropy completely.

There is an analytic solution to the field equations in the NLSM which
describes an unwinding knot. The field configuration is spherically
symmetric and remains on the vacuum manifold at all times. In
spherical polar coordinates $(r,\theta,\phi)$,
\begin{equation}
\Phi = \left(\cos\chi(r,t),
             \sin\chi(r,t)\,\cos\theta,
             \sin\chi(r,t)\,\sin\theta\,\cos\psi,
             \sin\chi(r,t)\,\sin\theta\,\sin\psi \right)
\end{equation}
has only one degree of freedom, $\chi(r,t)$, remaining, and any
non-singular configuration has $\chi(0, t) = 0$ or $\pi$. In this
form the SSSS solution is
\begin{equation}
\chi(t,r)=\left\{\begin{array}{ll}
                    2\tan^{-1}(-r/t)     & \quad t < 0     \\
	            2\tan^{-1}( t/r) + \pi & \quad 0 < t < r \\
                    2\tan^{-1}( r/t) + \pi & \quad 0 < r < t
	         \end{array} \right.
\end{equation}
At the origin $\chi(0, t)$ jumps discontinuously from $0$ to $\pi$ at
the unwinding at $t = 0$, and the outgoing solution has a gradient
discontinuity at $r = t$ (though the stress-energy tensor remains
smooth). Applying Eq.~(\ref{eAS1}) to the SSSS solution one reproduces
the result of Ref.~\cite{TEXMTS}, i.e.
\begin{equation}
\label{eSSSSDT}
\frac{\Delta T}{T} = \frac{t_{i}}
	{\sqrt{2|{\bf x}_\perp|^2 + t_{i}^2}}\;\epsilon
\end{equation}
where $\epsilon = 8\pi^{2}G\Phi_{\rm o}^{2}$, $|{\bf x}_\perp|$ gives
the impact parameter of the photon from the texture centre, and
$t_{i}$ gives the time that the photon sheet passes through the
centre. The pattern is a cold spot of depth $-\epsilon$ if the photons
pass the centre before unwinding ($t_{i}<0$) and a hot spot of height
$+\epsilon$ if the photons pass after unwinding ($t_{i}>0$).  The Full
Width Half Maximum (FWHM) of the hot/cold spot, that is, the diameter
of the circular ring centered on the spot, along which the temperature
is half the central value, is $\sqrt{6}\,|t_{i}|$ and grows unbounded
both before and after the unwinding.

\subsection{Numerical Methods}

For each initial field configuration under investigation the CMB
anisotropy is computed in much the same way as in Ref. \cite{BBS}. For
each simulation we follow planes of photons once across the simulation
cube, utilising the periodic boundary conditions to allow each plane
to travel the same distance through the simulation. The photon planes,
each separated by one lattice spacing, are taken to be orthogonal to
one of major axes of the cube. We may label each plane by its initial
co-ordinate along that axis which corresponds to the label $t_{i}$
defined above. Following the sheets in sequence then gives the
evolution of the anisotropy in time as viewed by the observer. On each
plane the temperature is calculated on a grid with the same spacing as
the grid for the texture evolution. At each gridpoint at each time
step the field's spatial and temporal derivatives are calculated and
the integrand of Eq.~(\ref{eAS3}) is summed over the run-time to
obtain ${\bf U}$. We then calculate $\Delta T/T$ from ${\bf U}$ using
Fast Fourier Transforms.

For configurations admitting an unwinding event 20 photon sheets are
chosen such that half pass the unwinding site before the unwinding
event occurs and half afterwards. In order to include the necessary
post-unwinding field evolution we only consider configurations known
to unwind within the first two-thirds of the total run time. Our
catalogue of random field configurations \cite{BCL2} contains 11 whose
unwinding event meets this criterion, and the specimen spherically
symmetric configuration is explicitly chosen so to do. To these we add
a further 11 non-unwinding configurations for which we follow all 96
photon sheets. We thus throw out the vast majority of random
configurations since only a few percent of such configurations exhibit
an unwinding.  Our sample therefore has a far greater proportion of
unwindings than for patterns picked at random.

\section{Results}

\subsection{CMB Amplitude}

As discussed above the SSSS unwinding has a constant CMB anisotropy
peak height $\Delta T/T = \pm \epsilon$. As a test of our evolution
and CMB anisotropy codes we are able to reproduce this analytic result
numerically to within 5\%. By comparison the more general spherically
symmetric configuration has peak heights
\begin{eqnarray}
\left. \frac{\Delta T}{T} \right|_{\rm min} & = &
\left\{
\begin{array}{ll}
-0.56 \;\epsilon & \hspace{1cm} {\rm XCORE} \\
-0.60 \;\epsilon & \hspace{1cm} {\rm NLSM}
\end{array}
\right.
\nonumber \\
\left. \frac{\Delta T}{T} \right|_{\rm max} & = &
\left\{
\begin{array}{ll}
+0.67 \;\epsilon & \hspace{1cm} {\rm XCORE} \\
+0.66 \;\epsilon & \hspace{1cm} {\rm NLSM}
\end{array}
\right.
\end{eqnarray}
Here and subsequently, the asymmetry between the maxima and minima is
presumably attributable to the minima being generated earlier in the
evolution (photons climbing out of the collapsing texture) than the
maxima (photons falling in with the collapsing texture); the field
correlations are therefore more pronounced for the maxima
systematically enhancing the anisotropy.

Texture field evolution generically includes two sources of metric
perturbation. The first is from the collapse, and possible unwinding,
of gradient energy on sub-horizon scales, and generates primarily
short wavelength modes. The second is from the correlation of the
field on horizon scales, and generates primarily long wavelength
modes. For spherically symmetric configurations the field is
artificially correlated on all scales at the outset, and the second
term is not present. However, for random initial conditions this is
not the case, and we expect contributions from both sources. The mean
peak height associated with the random texture configuration unwinding
events is
\begin{eqnarray}
\label{RNDDT}
\left. \frac{\Delta T}{T} \right|_{\rm min} & = &
\left\{
\begin{array}{ll}
(-0.62 \pm 0.17) \;\epsilon & \hspace{1cm} {\rm XCORE} \\
(-0.66 \pm 0.18) \;\epsilon & \hspace{1cm} {\rm NLSM}
\end{array}
\right.
\nonumber \\
\left. \frac{\Delta T}{T} \right|_{\rm max} & = &
\left\{
\begin{array}{ll}
(+0.76 \pm 0.17) \;\epsilon & \hspace{1cm} {\rm XCORE} \\
(+0.78 \pm 0.20) \;\epsilon & \hspace{1cm} {\rm NLSM}
\end{array}
\right.
\end{eqnarray}
where the error bars here and below are one standard deviation over
the set of simulations in question.

We are also interested in the typical amplitude of the background,
long wavelength, anisotropy modes. Reading the amplitudes directly
from the simulations with no unwinding event is problematic since even
here the gradient energy may locally collapse before being annihilated
by large scale correlation, introducing short wavelength modes. We
find significant non-unwinding events in all but two of the
configurations investigated, with over one third of them exhibiting
two events. However, none of these occur in events with a true
unwinding. The mean anisotropy across each sheet has been set to zero
but we can still quantify the amplitude of the long wavelength
fluctuations about this by calculating the mean modulus and the
standard deviation of the anisotropy averaged over the entire pattern
for all simulations. These quantities are measures of the
long-wavelength modes since the high peaks which contain most of
short-wavelength power subtend a small solid angle. Taking all 22
simulation datasets together we find
\begin{eqnarray}
\overline{\left|\frac{\Delta T}{T}\right|} & = & 0.12\;\epsilon
\nonumber \\
\sigma & = & 0.17\;\epsilon
\end{eqnarray}
in both the XCORE and NLSM approximations.

The non-unwinding events are of considerable interest, especially in
the light of the rarity of actual unwindings. The mean peak
anisotropies associated with them are found to be
\begin{eqnarray}
\label{NONDT}
\left. \frac{\Delta T}{T} \right|_{\rm min} & = &
\left\{
\begin{array}{ll}
(-0.65 \pm 0.29) \;\epsilon & \hspace{1cm} {\rm XCORE} \\
(-0.67 \pm 0.29) \;\epsilon & \hspace{1cm} {\rm NLSM}
\end{array}
\right.
\nonumber \\
\left. \frac{\Delta T}{T} \right|_{\rm max} & = &
\left\{
\begin{array}{ll}
(+0.60 \pm 0.25) \;\epsilon & \hspace{1cm} {\rm XCORE} \\
(+0.63 \pm 0.25) \;\epsilon & \hspace{1cm} {\rm NLSM}
\end{array}
\right.
\end{eqnarray}

In order to describe features common to both unwinding and
non-unwinding events we shall henceforth use the notion of the event
midtime, being the time at which the associated anisotropy changes
from being a cold to a hot spot; this is clearly identical to the
unwinding time, where one exists. Table 1 provides a quick comparison
of the maximum and minimum anisotropies from the different field
configurations and numerical methods.

\subsection{CMB Spot Size}

The size of the CMB spot can be parameterized by the radius at which
the anisotropy falls to some specified threshold, $|\Delta T/T| = \xi
\epsilon$, at any given time. From Eq.~(\ref{eSSSSDT}) for the SSSS
solution this is
\begin{equation}
R_{\xi}(t_{\rm i}) = \sqrt{\frac{1 - \xi^{2}}{2\xi^{2}}} \, |t_{\rm i}|
\end{equation}
For the more general spherically symmetric configuration we can simply
read off the value of the threshold radius from the numerical
calculations. In the case of the random configurations the spot radius
is calculated by counting all points in the vicinity of the peak with
$\Delta T/T \leq -\xi \epsilon$ on sheets whose impact time is before
the event's midtime, and with $\Delta T/T \geq +\xi \epsilon$ on those
for which it is afterwards, and taking
\begin{equation}
R_{\xi}(t_{\rm i}) = \sqrt{\frac{N_{\xi}(t_{\rm i})}{\pi}}
\end{equation}
where $N_{\xi}(t_{\rm i})$ is the number of lattice points satisfying
the appropriate criterion at any impact time. We take the threshold
value to be $\xi = 0.5 \;\epsilon$, being $3\sigma$ in the background
anisotropy determined above.

Figure 1 shows the variation in mean spot radius with time in each
case. In contrast to the linearly divergent SSSS case, in all the
other cases the spot radius reaches a maximum before falling off at
early and late times. Except at their midtimes, the random and SS
configurations' maximum radii are typically between 1/5 and 1/3 of
that of the SSSS solution at the same time.

We would now like to make some estimate of the angular size that these
spots would have on the microwave sky. This is never going to be
completely satisfactory while we restrict ourselves to flat space
simulations, but nevertheless estimates can be made. Recalling our
designation of the initial horizon radius as 32 grid units via an
estimate of the correlation length, and given that unwindings
typically occur after around 30 time units, the horizon radius at
unwinding is around 60 grid units. This is to be compared with a
typical threshold spot radius of around 3 grid units, so we estimate
the spot radius to be of the order of 5\% of the horizon radius.

Independent of any simulation parameters, we know directly from
cosmology the angle subtended by the horizon at last scattering. As a
function of the redshift $z$ of last scattering; it is simply
\begin{equation}
\theta_{{\rm hor}} \simeq (1 + z_{{\rm ls}})^{-1/2} \; {\rm radians}
\end{equation}
The angular radius of the spot on the microwave sky is hence of order
\begin{equation}
\label{eANG}
\theta_{\xi} \sim \frac{R_{\xi}}{R_{{\rm hor}}(t_{{\rm uw}})} \, \theta_{{\rm
hor}}
\end{equation}
where $R_{{\rm hor}}(t_{{\rm uw}})$ is the horizon size at the event's
midtime. In the spherically symmetric cases the correlation of the
field on super-horizon scales means that the CMB spot may become
larger than the horizon away from the unwinding event.  This is
certainly the case for the SSSS solution, where the spot size grows
linearly with impact time, although we note that we would only trust
the result Eq.~(\ref{eSSSSDT}) close to the unwinding site. By
contrast the general spherically symmetric configuration spot size
shows similar behaviour to those of the random configurations,
reaching a maximum radius both before and after the midtime. In the
SSSS case, since a super-horizon sized coherent anisotropy is
unphysical we impose a cutoff at the horizon size, setting the
prefactor in Eq.~(\ref{eANG}) to one, though this is still a far from
satisfactory state of affairs suggesting that, at least in estimating
metric perturbations in the texture model, one should steer well clear
of the SSSS solution whenever possible.

For textures unwinding at decoupling, assuming that these results can
be carried into the matter dominated epoch and that the redshift of
last scattering is given by one of
\begin{equation}
1+z_{{\rm ls}} \sim \left\{ \begin{array}{ll}
			1000 & \mbox{no re-ionisation} \\
			50 & \mbox{re-ionisation}
		      \end{array} \right.
\end{equation}
this yields
\begin{equation}
\theta_{\xi} \sim \left\{ \begin{array}{ll}
			0.1^{\circ} & \mbox{no re-ionisation} \\
			0.4^{\circ} & \mbox{re-ionisation}
		   \end{array} \right.
\end{equation}
Bearing in mind that these are radii, our predicted spot-size appears
similar to that found in Ref.~\cite{CPT} who only considered a
cosmology with reionisation.

Illustrating these features together, Figures 2a--d shows the spot
anisotropy profile for the SSSS solution, the spherically symmetric
configuration, a typical unwinding event and a typical non-unwinding
event. In each case the profiles are taken immediately before, at, and
immediately after the midtime.

\section{Discussion}

It is encouraging to note the strong consistency between the results
in the XCORE and NLSM approximations. Not only can the statistical
results over the ensemble of simulations be treated as identical, but
individual simulations with the same initial conditions exhibit events
whose individual properties are quantitatively extremely similar.
Since the two approximation methods are very different, this
congruence of results lends considerable weight to our belief that
both sets of simulations are accurately modelling the texture field's
evolution. While this is only the case at the level of resolution
employed here (in both approximations any significant reduction in
resolution is seen to introduce serious numerical errors --- see
Ref.~\cite{B94} for details), we are confident that the results in
this paper are trustworthy and allow a genuine comparison between
numerical calculations and the analytic solutions.

Our key results from this comparative study are the following
\begin{enumerate}
\item The randomly generated configurations produce microwave
anisotropy patterns with completely different properties to the exact
NLSM solution.
\item Localised concentrations of gradient energy which do {\em not}
lead to unwindings can still generate anisotropies which are extremely
similar to those generated by genuine unwindings.
\item We find a characteristic spot size considerably smaller than the
horizon size at the collapse time.
\end{enumerate}
Let us comment on each of these in turn, before making some final
comments on the relation between our work and that of other authors.

We have found considerable differences between the properties of the
microwave aniso\-tropies generated by the exact SSSS solution and our
more realistic randomly generated configurations. The peak anisotropy
of the random configurations is smaller by 20--40\%, and much more
significantly the spot size is considerably smaller, leading to a huge
reduction in the spot area on the sky and hence in the anisotropy
integrated over a beam profile. This effect can doubtless be
attributed to the SSSS configuration's spherical symmetry imposing
unphysical correlations on scales vastly greater then the horizon size
whilst its self-similarity forces the collapse to occur as fast as
causally possible, together leading to a maximisation of the
anisotropies. The failure of the SSSS solution to represent the
properties of true textures appears to be even more dramatic when
looking at microwave anisotropies than the failure already noted
\cite{BCL1,BCL2} when only considering the unwinding dynamics.

Perhaps our most significant result is the observation that localised
concentrations of gradient energy which nevertheless fail to unwind
can still lead to very substantial microwave anisotropies; indeed the
anisotropies due to such non-unwinding events are indistinguishable
from those due to unwinding events. Furthermore the number density of
such events is vastly greater than that of unwinding events; unwinding
events occur only in around 4\% of our simulations whilst
non-unwinding events typically occur once per simulation. Given their
indistinguishability we are led to believe that the non-unwinding
events are the dominant contributors to the microwave background
anisotropies, with unwinding events playing only a minor role.

Our estimate of the spot size is that their typical radius is
around 5\% of the horizon radius at the time the spot is generated.
As stated above, this is considerably smaller than that generated by
the SSSS solution, although that solution can only have a spot size
attributed at all by the imposition of a horizon-size cut-off to
remove the unphysical correlations. If we follow most authors in
assuming that the texture scenario re-ionises the universe to a
redshift of around 50, and take the liberty of extrapolating our flat
space results directly into a matter dominated universe, then we
anticipate a typical spot diameter, be it generated by an unwinding or
a non-unwinding event, of about a degree. By contrast, in the absence
of re-ionisation the typical spot diameter is only of the order of
$10$ arcminutes.

Our study has been entirely a comparative one, concentrating on the
relative results from alternative choices of initial conditions and
approximation methods in the simplest possible environment. The
emphasis has been primarily focussed on the behaviour of individual
textures. We have not attempted any full cosmological simulations, and
indeed the computing resources which we believe are required to mount
accurate simulations of ensembles of textures are beyond those
available at present (see Refs.~\cite{B93,B94} for a detailed
discussion). The cosmological simulations that have been done
\cite{PST,CPT,BR} may therefore include systematic errors due to
their inability sufficiently to resolve the microphysical events
associated with the concentration of field gradient energy. In
particular it should be noted that low resolution simulations are
susceptible to spurious unwinding events, where configurations
covering less than half of the vacuum manifold nonetheless unwind.
With this caveat firmly in mind we now compare our results with those
from such simulations.

The most interesting paper to consider in this context is that of
Coulson {\it et al} \cite{CPT}, who include a large number of physical
processes in their cosmological NLSM simulations in an attempt to
generate predictions for degree scale anisotropies. We are unable to
say anything about their quantitative value for the magnitude of the
anisotropy, or even about the statistics, as we have no guidance as to
how to combine our different choices of initial conditions into an
ensemble. Nevertheless, in two important {\it qualitative} aspects we
are in good agreement with their results. Firstly, their spot size is
typically of order one to two degrees (it is slightly hard to estimate
which as they smooth their data with a gaussian filter on around this
scale). This is considerably less than the horizon size as they assume
reionisation, and hence is in conflict with the original expectation
based on the NLSM solution that the spots would be of order ten
degrees \cite{TEXMTS,DS}. However it is consistent with our result
that typical texture unwindings do not lead to such large spots.
Secondly, they exhibit a diagram of their simulation covering an area
thirty degrees on a side (i.e.~about 2\% of the sky) which includes
several spots. However, the number density of texture unwindings is
known to be so low \cite{TEX,B93} that it is very unlikely that even a
single unwinding would have occurred in such an area. It is possible
therefore that their spots actually correspond to non-unwinding
events, which we have seen greatly dominate over unwinding events
while giving rise to a very similar signature, although such results
would also be consistent with spurious unwinding events. Qualitatively
therefore our picture of many small spots is consistent with their
results, although we emphasise again that we have no means of
assessing their quantitative estimates.

However, our results do seriously undermine the alternative method of
estimating microwave background anisotropies, where photons are traced
through an ensemble of SSSS initial configurations which are evolved
in flat or matter-dominated backgrounds \cite{DS}. The spots found in
such calculations are probably too bright by a modest factor, but much
more significantly the area is overestimated by a factor which could
easily be in the range 25--100 --- both factors being attributable to
the use of a maximal coherent initial velocity field. On the other
hand such simulations only include unwinding events, whereas in
practice the all-sky anisotropy is dominated by non-unwinding events,
so the number of spots appearing in their simulations is down,
conceivably by a similar factor. Such opposing effects may enable
simple measures such as the all-sky rms fluctuation to emerge in
acceptable agreement with other derivations (e.~g.~Ref.~\cite{PST}),
despite the detailed
microwave anisotropy distribution being totally different.

\section*{Acknowledgements}

JB and EJC are supported by the SERC, ARL by the SERC and the Royal Society, AS
by the DOE and NASA grant NAGW-2381 at Fermilab and SVR by the NRC. EJC, ARL,
AS and SVR would like to thank the Aspen Center for Physics for their
hospitality while part of this work was done. We thank Mark Hindmarsh for
helpful discussions. ARL acknowledges the use of the
STARLINK computer system at the University of Sussex.

\frenchspacing


\newpage
\vspace{24pt}
\begin{tabular}
   {|l|c|c|}\hline
    Configuration/Method\quad
     & ${\rm \Delta T/T}_{\rm max}$ \quad
      & ${\rm \Delta T/T}_{\rm min}$ \quad \\ \hline
    SSSS/Analytic & +1         & -1            \\ \hline
    SS/XCORE      & 0.67       & -0.56         \\ \hline
    SS/NLSM       & 0.66       & -0.60         \\ \hline
    Random/XCORE  & 0.76$\pm$0.17& -0.62$\pm$0.17  \\ \hline
    Random/NLSM   & 0.78$\pm$0.20& -0.66$\pm$0.18  \\ \hline
    Non-unwinding/XCORE  & 0.60$\pm$0.25  & -0.65$\pm$0.29   \\ \hline
	Non-unwinding/NLSM   & 0.63$\pm$0.25  & -0.67$\pm$0.29   \\ \hline
\end{tabular}

\vspace{24pt}
\noindent
{\em Table 1}\\
Comparison of the peak temperature anisotropies as
described in the text. Here SS is the spherically symmetric
configuration of Eq. (\ref{SS}) in \S2.1.  The anisotropies are in
units of $\epsilon=8\pi^2 G\Phi_{\rm o}^{2}$.

\vspace{24pt}

\section*{Figure Captions}

\vspace{24pt}
\noindent
{\em Figure 1}\\
The radii of microwave spots above a given anisotropy threshold, as described
in
the text. The solid line corresponds to the exact SSSS solution, the dot-dashed
to the spherically symmetric simulation, the dashed to the average of the
unwinding events and the dotted to the average of the non-unwinding events.

\vspace{24pt}
\noindent
{\em Figures 2a--d}\\
The fractional anisotropy generated by the evolution of various initial field
configurations in units of $\epsilon$. These are (a) the SSSS exact solution,
(b) the spherically symmetric configuration, (c) a typical random unwinding
configuration, and (d) a typical random non-unwinding configuration. In each
case three photon sheets are shown, being three gridtimes before, at, and three
gridtimes after the event midtime.


\begin{thebibliography}{99}
\bibitem{INFL} E. W. Kolb and M. S. Turner, {\em The Early Universe},
	Addison-Wesley (1990).\\ A. R. Liddle and D. H. Lyth, Phys. Rep. 231,
	1 (1993).
\bibitem{STRING} T. W. B. Kibble, J. Phys. {\bf A9}, 1387 (1976).
\bibitem{TEX} N. Turok, Phys. Rev. Lett. {\bf 63}, 2625 (1989).
\bibitem{GS} A. Vilenkin, Phys. Rep. 121, 263 (1985).
\bibitem{BR} D. P. Bennett and S. H. Rhie, Phys. Rev. Lett. {\bf 65}, 1709
	(1990).
\bibitem{TEXLSS} R. Y. Cen, J. P. Ostriker, D. Spergel and N. Turok,
	Astrophys. J. {\bf 383}, 1 (1991).\\ A. K. Gooding, C. Park, D.
	Spergel, N. Turok and J. R. Gott, Astrophys J {\bf 393}, 42 (1992).
\bibitem{TEXMTS} N. Turok and D. N. Spergel, Phys. Rev. Lett. {\bf 64}, 2736
	(1990).
\bibitem{DS} R. Durrer and D. N. Spergel, in ``Trends in Astroparticle
	Physics'' ed. D Cline \& R. Peccei, World Scientific (Singapore)
	(1992).\\
	R. Durrer, A. Howard and Z-H Zhou, Phys. Rev. D{\bf 49}, 681
	(1994).
\bibitem{BR2} D. P. Bennett and S. H. Rhie, Astrophys. J. Lett. {\bf 406},
        L7 (1993).
\bibitem{PST} U.-L. Pen, D. N. Spergel and N. Turok, Phys. Rev. D{\bf 49},
692
	(1994).
\bibitem{CPT} D. Coulson, P. Ferreira, P. Graham and N. Turok, ``Microwave
	Anisotropies from Cosmic Defects'', Nature, in press (1994).
\bibitem{COBE} G. F. Smoot {\it et al}, Astrophys. J. Lett. {\bf 396}, L1
	(1992).
\bibitem{BCL1} J. Borrill, E. J. Copeland and A. R. Liddle, Phys. Rev. D{\bf
	46}, 524 (1992).
\bibitem{BCL2} J. Borrill, E. J. Copeland and A. R. Liddle, Phys. Rev. D{\bf
	47}, 4292 (1993).
\bibitem{NSY} M. Nagasawa, K. Sato and J. Yokoyama, Publ. Astron. Soc. Japan
        {\bf 45}, 755 (1993).
\bibitem{VS90} S. Veeraraghavan and A. Stebbins, Astrophys. J. {\bf 365}, 37
	(1990).
\bibitem{VS92SV93a} S. Veeraraghavan and A. Stebbins, Astrophys. J. Lett.
	{\bf 395}, L55 (1992). \\ A. Stebbins and S. Veeraraghavan,
	Phys. Rev. D{\bf 48}, 2421 (1993).
\bibitem{B94} J. Borrill, ``Numerical Methods in Cosmological Global Texture
	Simulations'', Imperial College preprint IMPERIAL/TP/93-94/22 (1994).
\bibitem{PFTV} W. H. Press, B. P. Flannery, S. A. Teukolsky and W. T.
        Vetterling, {\it Numerical Recipes} (Cambridge University Press 1986)
\bibitem{PRS} W. H. Press, B. S. Ryden and D. N. Spergel, Astrophys. J.
        {\bf 347}, 590 (1989).
\bibitem{STPR} D. N. Spergel, N. Turok, W. H. Press and B. S. Ryden, Phys.
	Rev. D{\bf 43}, 1038 (1991).
\bibitem{B93} J. Borrill, Phys. Rev. D{\bf 47}, 4298 (1993).
\bibitem{SW} R. Sachs and A. Wolfe, Astrophys. J. {\bf 147}, 73 (1967).
\bibitem{S88} A. Stebbins, Astrophys. J., {\bf 327}, 584 (1988).
\bibitem{H93} M. Hindmarsh, ``Small Scale Microwave Background Fluctuations
	from Cosmic Strings'', Astrophys. J., in press (1994).
\bibitem{SV93b} S. Veeraraghavan and A. Stebbins, ``Beyond the Small
Angle
	Approximation for MBR Anisotropy from Seeds'', Fermilab
preprint
	Fermilab-Pub-94/047-A (1994).
\bibitem{VS93} S. Veeraraghavan and A. Stebbins, in preparation (1994).
\bibitem{BBS}  F. Bouchet, D. Bennett and A. Stebbins, Nature {\bf 335}, 410
	(1989).
\end{thebibliography}
\end{document}